\documentclass[a4paper]{article}

\usepackage{INTERSPEECH2019}
\usepackage[dvipsnames]{xcolor}
\usepackage{colortbl,booktabs}
\usepackage{multicol}
\usepackage{enumitem}
\usepackage{mathtools}
\usepackage{algorithm}
\usepackage{algorithmic}
\newcommand{\etal}{\textit{et al}. }
\DeclareMathOperator*{\argmin}{arg\,min}

\usepackage{hyperref}
\hypersetup{
    colorlinks=true,
    linkcolor=blue,
    filecolor=magenta,      
    urlcolor=cyan,
    pdftitle={Sharelatex Example},
    bookmarks=true,
    pdfpagemode=FullScreen,
    }
\title{AutoSpeech: Neural Architecture Search for Speaker Recognition}
\name{Shaojin Ding$^1$\textsuperscript{*}, Tianlong Chen$^2$\textsuperscript{*}, Xinyu Gong$^{1,2}$, Weiwei Zha$^3$, Zhangyang Wang$^2$\thanks{* Equal contribution.}}
\address{
  $^1$Department of Computer Science and Engineering, Texas A\&M University, USA\\
  $^2$Department of Electrical and Computer Engineering, The University of Texas at Austin, USA\\
  $^3$School of Software Engineering, University of Science and Technology of China, China}
\email{shjd@tamu.edu, \{tianlong.chen, xinyu.gong, atlaswang\}@utexas.edu, Sa517009@mail.ustc.edu.cn}

\begin{document}

\maketitle
\begin{abstract}
Speaker recognition systems based on Convolutional Neural Networks (CNNs) are often built with off-the-shelf backbones such as VGG-Net or ResNet. However, these backbones were originally proposed for image classification, and therefore may not be naturally fit for speaker recognition. Due to the prohibitive complexity of manually exploring the design space, we propose the first neural architecture search approach for the speaker recognition tasks, named as \textbf{AutoSpeech}. Our algorithm first identifies the optimal operation combination in a neural cell and then derives a CNN model by stacking the neural cell for multiple times. The final speaker recognition model can be obtained by training the derived CNN model through the standard scheme. To evaluate the proposed approach, we conduct experiments on both speaker identification and speaker verification tasks using the VoxCeleb1 dataset. Results demonstrate that the derived CNN architectures from the proposed approach significantly outperform current speaker recognition systems based on VGG-M, ResNet-18, and ResNet-34 backbones, while enjoying lower model complexity.
\end{abstract}


\section{Introduction}
Speaker recognition aims to retrieve the identity of the speaker given his/her utterances. According to the content similarity of the utterances, speaker recognition can be categorized into text-dependent and text-independent, the latter being more general and realistic for practical applications. Additionally, according to different application settings, speaker recognition usually fall into one of the two categories: speaker identification (SID) and speaker verification (SV). SID identifies the speaker of an utterance from a known speaker set, and SV determines if the speaker of an utterance matches its given enrollment.

End-to-end speaker recognition systems \cite{variani2014deep, wan2018generalized, xie2019utterance, sadjadi2016ibm, li2017deep, snyder2018x} have emerged in recent years and achieved the state-of-the-art performance. These models usually admit a \textit{three-stage pipeline}: (1) A deep neural network, typically Convolutional Neural Networks (CNNs) or Recurrent Neural Networks (RNNs), as a feature extractor to generate frame-wise speaker embedding; (2) a temporal aggregation layer, to produce fixed length speaker embedding (i.e., d-vector); (3) A loss function to optimize the entire network. In testing stage, the network along with the temporal aggregation layer are first used to produce d-vectors for the testing utterances, and then a similarity metric (e.g., cosine similarity) follows to generate the final same/different speaker decision based on the d-vectors.

Most CNN/RNN-based speaker recognition works focus on improving the effectiveness of the speaker embedding through advanced training objectives and temporal aggregation strategies \cite{gao2019improving, hajibabaei2018unified, xie2019utterance, cai2018exploring, li2017deep}. By contrast, network architecture design has received relatively less attention. Existing studies usually use off-the-shelf backbones that have been shown to be successful in standard tasks such as image classification (e.g., VGG-Net \cite{simonyan2014very}, ResNet \cite{he2016deep}). However, these backbones are not designed for and may not be optimal for speaker recognition. Meanwhile, in other speech processing tasks such as speech recognition and speech synthesis, the improved design of network architecture has already yielded evident performance gains \cite{chan2016listen, shen2018natural}. In view of those, we conjecture that \textbf{enhancing network architecture design matters for improving speaker recognition performance too}. Yet as the major hurdle for improving the backbone of any new task, manually exploring the gigantic design space of deep networks is notoriously tedious and ad-hoc. 


This paper proposes an automated approach to identify the optimal CNN architecture for text-independent speaker recognition, named as \textbf{AutoSpeech}. It is inspired by recent advances on neural architecture search (NAS) \cite{liu2018darts, real2019regularized, pham2018efficient, zoph2016neural, negrinho2017deeparchitect}, which has proven success in designing deep networks that outperform hand-designed best performers in various tasks \cite{baruwa2019leveraging, liu2019auto, gong2019autogan, quan2019auto}. 
We evaluated the effectiveness of AutoSpeech on both SID and SV tasks using VoxCeleb1 dataset \cite{nagrani2017voxceleb}. Our derived CNN architectures significantly outperforms several networks used in state-of-the-art speaker recognition systems (based on VGG-M, ResNet-18, and ResNet-34), at lower model complexities.


\section{Literature Review}
\label{sec:literature}

\subsection{Speaker Recognition}
Earlier speaker recognition systems used spectral features (e.g., MFCCs) as the speaker embedding \cite{martinez2012speaker, tiwari2010mfcc}, but these systems have been superseded by systems based on “i-vectors” \cite{dehak2010front, garcia2011analysis, kenny2010bayesian}. An i-vector takes speech from a speaker and uses it to adapt a speaker-independent Gaussian Mixture Model (GMM, referred to as a Universal Background Model). The means of the adapted GMM are concatenated to form a supervector, which is then reduced in dimensionality using joint factor analysis. More recently, end-to-end speaker recognition \cite{variani2014deep, wan2018generalized, xie2019utterance, sadjadi2016ibm, li2017deep} systems have been shown to surpass the i-vector based systems and achieve state-of-the-art performance. Variani \etal first proposed a neural network based speaker recognition system. In their work, they used maxout fully-connected network to produce d-vectors, and then they used cosine similarity to make the final decision. Following this, advanced network architectures such as CNNs \cite{xie2019utterance, li2017deep, cai2018exploring, hajibabaei2018unified, bhattacharya2019deep} and RNNs \cite{li2017deep, wan2018generalized} are used for feature extracts. Additionally, advanced training objectives \cite{wan2018generalized, hajibabaei2018unified, bhattacharya2019deep} and temporal aggregation strategies \cite{gao2019improving, xie2019utterance, cai2018exploring, li2017deep} are also proposed to improve speaker recognition performance.

\subsection{Neural Architecture Search}
The goal of neural architecture search is to search for an optimal network architecture for a given task. To accomplish this, a search space will be constructed firstly, which may vary across different tasks and objectives. Towards accurate image classification, many previous work \cite{liu2018darts,real2019regularized,pham2018efficient} adopt a cell-based search space with complicated inner connection. To improve its efficiency, \cite{tan2019mnasnet,wu2019fbnet,dai2019chamnet} adopts a sequential-based search space, benefiting a lot in reducing latency and FLOPs inherently. More recently, neural architecture search has also been applied to other tasks (e.g., image segmentation \cite{liu2019auto, chen2019fasterseg}, generative adversarial network \cite{gong2019autogan}, and phone recognition \cite{baruwa2019leveraging}), where task-specific search spaces are usually designed. To conduct architecture search over the defined search space, various optimization methods are proposed. Barret and Quoc first use the reinforcement learning (RL) to search for architecture\cite{zoph2016neural}, where a RNN serves as an architecture sampler. However, it is quite time-consuming (2,000 GPU days). To address this limitation, Hieu \etal propose to use weight sharing (ENAS\cite{pham2018efficient}) to accelerate RL-based search method. Furthermore, Liu \etal present a gradient-based method \cite{liu2018darts} to accelerate the searching process.


\section{Methods}
\label{sec:methods}

\begin{figure}
	\centering
	\includegraphics[width=\columnwidth]{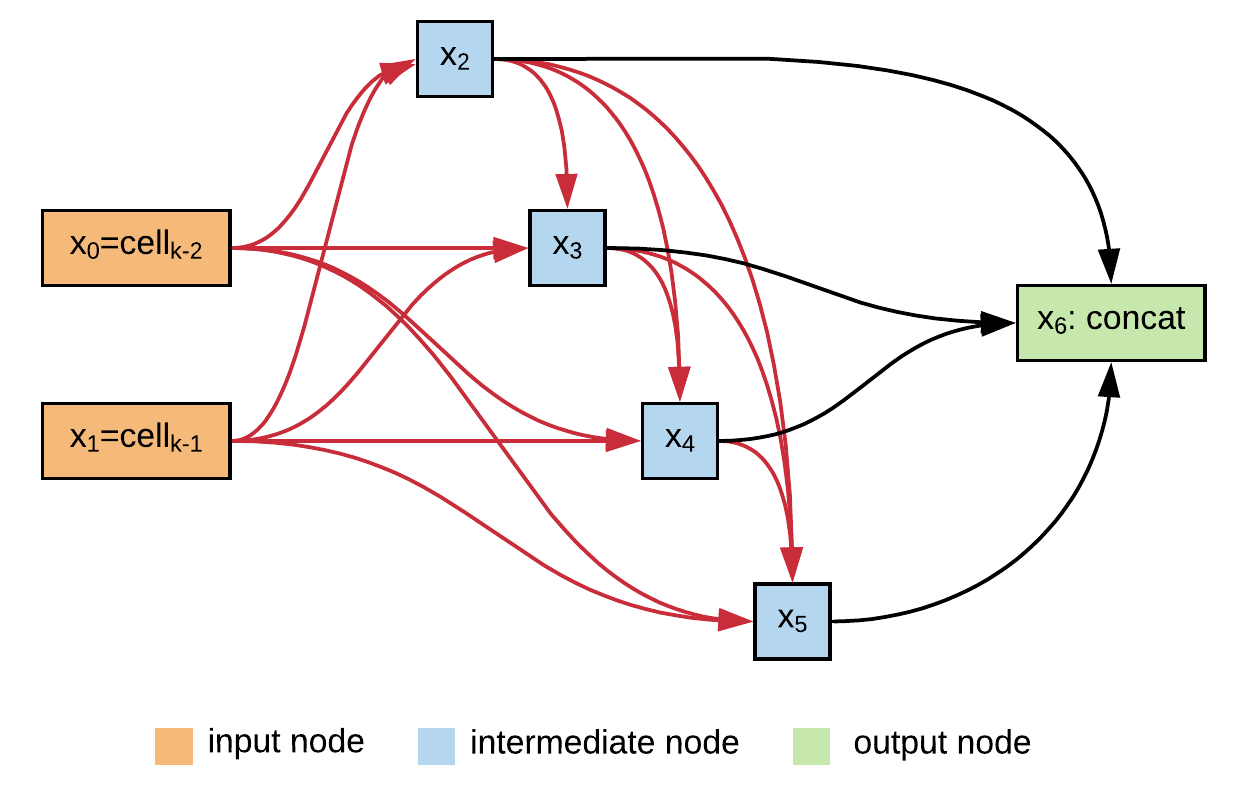}
	\vspace{-1em}
	\caption{Illustration of a neural cell. During the search process, the intermediate nodes ($x_2$ to $x_5$) are densely connected (red arrows), and the optimal combination of the operations on these edges are found at the end of the searching process. During architecture derivation, only the two operations with highest softmax probabilities (excepting zero operation) are retained for each intermediate node.}
	\vspace{-1em}
	\label{fig:neural_cell}
\end{figure}

In this section, we introduce how to use NAS to automatically find the optimal CNN architecture for speaker recognition. Following previous NAS methods \cite{liu2018darts, zoph2016neural}, we use a \textit{two-stage pipeline}. First, we search for the optimal architecture of a neural cell that consists of several different types of operations. Second, we derive a CNN model using the searched cell, and then train the CNN model through the standard speaker recognition training scheme. We first define the search space of a neural cell in Section \ref{sec:search_space}. Then, we describe the NAS algorithm in Section \ref{sec:nas_algorithm}. Finally, we introduce how to derive a CNN model using searched neural cells and how to train it, in Section \ref{sec:derive}.

\subsection{Neural cell and search space}
\label{sec:search_space}

During the NAS process, the network is constructed by stacking multiple neural cells. We define a neural cell to have $N=7$ nodes, as illustrated in Figure \ref{fig:neural_cell}. A cell can be viewed as a directed acyclic graph, where a node $x_i$ corresponds to a tensor and a directed edge $(i, j)$ corresponds to an operation $o_{ij}(\cdot)$  Following \cite{liu2018darts, zoph2016neural, real2019regularized, liu2018progressive}, we set each cell to have two input nodes,  four intermediate nodes, and one output node. The input nodes are set equal to the outputs of previous two cells, respectively (e.g., The first node $x_0$ in the $k$-th cell is equal to the output of the ($k$-2)-th cell, and the second node $x_1$ in the $k$-th cell is equal to the output of the ($k$-1)-th cell). The four intermediate nodes ($x_2$ to $x_5$) are densely connected (red arrows in Figure \ref{fig:neural_cell}, 14 edges in total), and each of them is computed as the summation based on all of its predecessors:

\begin{equation}
    x_j \coloneqq \sum_{i<j}o_{ij}(x_i)
\label{eq:intermediate_node}
\end{equation}

\noindent
A best combination of the operations on these edges are found at the end of the searching process. Finally, the output node $x_7$ is defined as the concatenation of all intermediate nodes.

The search space $O$ defines all possible candidate operations in a neural cell. We include the following 8 operations that are prevalent in modern CNNs in our search space, and the resulting search space size is $8^14$ (14 edges, 8 candidate operations for each edge).
\vspace{-1em}
\setlength\columnsep{0pt}
\begin{multicols}{2}
\begin{itemize}
\advance\leftskip-10pt
    \item $3\times3$ separable convolution
    \item $5\times5$ separable convolution
    \item $3\times3$ dilated convolution
    \item $5\times5$ dilated convolution
    \item $3\times3$ average pooling
    \item $3\times3$ max pooling
    \item skip connection
    \item no connection (zero)
\end{itemize}
\end{multicols}
\vspace{-1em}
We stack the neural cell 8 times to form the backbone CNN during the NAS process. Additionally, we define two types of neural cells: \textit{normal cells} (cells that keep the spatial resolution of the feature tensor) and \textit{reduction cells} (cells that divide the spatial resolution by 2 and multiply the number of filters by 2). Following previous studies \cite{zoph2018learning, liu2018darts, real2019regularized}, we set the cells whose position is located at the $1/3$ and $2/3$ of the total depth to reduction cells, and the other cells are normal cells. All the normal cells share the same architecture, and all the reduction cells share the same architecture, respectively. The output of the last cell is then fed to an average pooling layer, followed by a fully-connected layer that outputs the softmax probability.


\begin{algorithm}[t]
\caption{Neural architecture search algorithm}
\label{alg:nas}
\begin{algorithmic}
\STATE \textbf{Inputs}: Training set $D_{train}$, validation set $D_{val}$.
\STATE \textbf{Output}: Searched neural cell architecture.
\\\hrulefill
\WHILE{not converge}
    \STATE - Update weight parameters $\omega$ by $\nabla_{\omega}L_{train}(\omega, \alpha)$ on $D_{train}$.
    \STATE - Update architecture parameters $\alpha$ by $\nabla_{\alpha}L_{val}(\omega, \alpha)$ on $D_{val}$.
\ENDWHILE
\STATE - Derive the neural cell architecture from $\alpha$ by retaining the two operations with highest softmax probabilities for each node.
\end{algorithmic}
\end{algorithm}

\subsection{Neural architecture search}
\label{sec:nas_algorithm}

To search for the optimal operation combinations in a neural cell, we define two sets of parameters: (1) a set of \textit{architecture} parameters $\alpha$ that controls the choice of operations and (2) a set of \textit{weight} parameters $\omega$ of all operations in $O$. We use architecture parameters $\alpha_{ij} \in \mathbb{R}^{|O|}$ to relax the categorical choice of a particular operation $o_{ij}$ on edge $(i, j)$ to a softmax over all possible operations in the search space:

\begin{equation}
    \hat{o}_{ij}(x_i) \coloneqq \sum_{o\in O}\frac{\exp(\alpha^o_{ij})}{\sum_{o'\in O}\exp(\alpha^{o'}_{ij})}o(x_i)
\label{eq:relaxation}
\end{equation}

\noindent
As a result, the search space becomes continuous, and NAS can be realized by optimizing the architecture parameters. In addition, since we have two types of neural cells (normal cells and reduction cells), the architecture parameters becomes $\alpha=(\alpha_{normal}, \alpha_{reduction})$, where $\alpha_{normal}$ are shared among all normal cells and $\alpha_{reduction}$ shared among all reduction cells.

We use a differentiable NAS algorithm \cite{liu2018darts} to learn $\alpha$ and $\omega$ jointly through back-propagation. We denote the training loss as $L_{train}(\omega, \alpha)$ and the validation loss as $L_{val}(\omega, \alpha)$. The NAS process can be viewed as a bi-level optimization problem \cite{colson2007overview} (eq. \ref{eq:bilevel}), which aims to find an optimal $\alpha$ that minimizes $L_{validation}(\omega, \alpha)$, where the optimal $\omega$ is determined by minimizing $L_{train}(\omega, \alpha)$.

\begin{align}
\label{eq:bilevel}
    & \min_{\alpha} L_{val}(\omega, \alpha) \\
    & \mathrm{s.t.} \;  \omega = \argmin_{\omega} L_{train}(\omega, \alpha) \nonumber
\end{align}

\noindent
In our task, we use cross-entropy loss for both $L_{train}$ and $L_{val}$:

\begin{equation}
\label{eq:loss}
    L \coloneqq - \sum_{k=1}^{K}\mathbb{I}(y=k)\log p_k
\end{equation}

\noindent
where $\mathbb{I}(\cdot)$ is the indicator function, $K$ is the number of speakers, $y$ is the ground-truth speaker, and $p_k$ is the softmax probability of speaker $k$. 

In practice, this bi-level optimization problem can be solved iteratively, as shown in Algorithm \ref{alg:nas}. During each iteration, we perform two steps: weight parameters update and architecture parameters update. In the weight parameters update step, we update $\omega$ by minimizing the the cross-entropy loss on training set $L_{train}$, while fixing $\alpha$. In the architecture parameter update step, we update $\alpha$ by minimizing the cross-entropy loss on validation set $L_{val}$, while fixing $\omega$.

The algorithm terminates when the choice of operations in the neural cell converges, which is empirically measured by the entropy of the architecture parameters $\alpha$ as:

\begin{equation}
\label{eq:entropy}
    E = \sum_{i, j} \sum_{o\in O} \alpha^o_{ij} \log \alpha^o_{ij}
\end{equation}

\noindent
A smaller the entropy indicates a higher confidence of choosing a particular operation among all possible operations. As a result, we consider the architecture of the neural cell being converged when the entropy does not decrease.


\subsection{Architecture derivation}
\label{sec:derive}

Once the search process is converged, we can derive a neural cell architecture from the learned architecture parameters. For each node $x_j$, we retain \textit{two operations with highest softmax probabilities} (excepting zero operation) from all its predecessors ${x_i}, \forall i<j$, following \cite{liu2018darts}. The softmax probability of an operation $o$ between nodes $(i, j)$ is defined as:

\begin{equation}
    p_{ij}^o \coloneqq \frac{\exp(\alpha^o_{ij})}{\sum_{o'\in O}\exp(\alpha^{o'}_{ij})}
\label{eq:relaxation}
\end{equation}

\noindent
With the derived neural cell, we construct a CNN by stacking multiple neural cells. We explore different number of neural cells and different number of channels in our experiments, as we will describe in Section \ref{sec:setup}. Similar to the CNN architecture in NAS process, we still set the cells whose position is located at the $1/3$ and $2/3$ of the total depth to reduction cells, and the other cells are normal cells. 

To obtain the final speaker recognition model, we train the CNN model from random initialized weights (denoted as \textit{training from scratch})  on training set by minimizing the cross-entropy loss in eq. \ref{eq:loss}. We use the output of the average pooling layer of the trained CNN as the speaker embedding.

\section{Experiments}
\label{sec:experiments}


\begin{table*}[ht]
\vspace{-0.5em}
\begin{center}
\caption{Speaker identification and speaker verification performance on VoxCeleb1 dataset. Dimensions indicate the dimensionality of the speaker embedding. $N$ denotes the number of neural cells, and $C$ denotes the number of initial channels.}
\vspace{-0.5em}
\label{table:results}
\resizebox{0.75\textwidth}{!}{
\begin{tabular}{c|cc|c|c|c}
\hline
Method &  Top-1(\%) & Top-5(\%) & EER (\%) & Dimensions & Parameters \\
\hline
VGG-M \cite{nagrani2017voxceleb} & 80.50 & 92.1 & 10.20 & 1,024 & 67 million \\
ResNet-18 \cite{hajibabaei2018unified, bhattacharya2019deep} & 79.48 & 90.97 & 12.30 & 512 & 12 million \\
ResNet-34 \cite{chung2018voxceleb2, xie2019utterance, cai2018exploring} &  81.34 & 94.49 & 11.99 & 512 & 22 million \\
\hline
Proposed ($N=8, C=64$) & 84.45 & 94.74 & 9.13 & 1,024 & 5 million \\
Proposed ($N=30, C=64$) & 83.45 & 94.21 & 9.01 & 1,024 & 18 million \\
Proposed ($N=8, C=128$) & \textbf{87.66} & \textbf{96.01} & \textbf{8.95} & 2,048 & 18 million \\
\hline
\end{tabular}}
\end{center}
\vspace{-20pt}
\end{table*}

\subsection{Datasets}
We conducted experiments on VoxCeleb1 \cite{nagrani2017voxceleb} dataset consisting of 153,516 utterances produced by 1,251 speakers. VoxCeleb1 has an identification split and a verification split for SID and SV, respectively. The identification split divides the entire dataset to a training set of 138,316 utterances, a development set of 6,904 utterances, and a test set of 8,251 utterances.  All the three sets were produced by the same 1,251 speakers. The verification split divides the dataset into a training set of 148,642 utterances produced by 1,211 speakers and a test set of 4,874 utterances produced by 40 speakers. The test speakers do not overlap with training. The verification split also provided 37,720 trial pairs for SV evaluation. During the NAS process, we followed the identification split. These subsets were used as introduced in Section \ref{sec:nas_algorithm}. During training from scratch, we train different models for SID and SV tasks following identification split and verification split, respectively.

\subsection{Implementation details}
For each utterance, we extracted a 257-dim spectrogram with a 25ms window and 10ms shift. We performed mean and variance normalization on each frequency bin of the spectrogram. To form a mini-batch, we randomly selected 3-second segments from utterances. As a result, the input size is $257\times 300$. 

We implemented the proposed method based on PyTorch \cite{paszke2017automatic} and NVIDIA TITAN RTX GPU\footnote{Codes available at \url{https://github.com/TAMU-VITA/AutoSpeech}}. During the NAS process, the network with 8 neural cells was training using the algorithm in Section \ref{sec:nas_algorithm} for 50 epochs with a batch size of 16. We set the initial channels to 16, due to the limited GPU memory. We used Adam Optimizer to optimize both the architecture parameters $\alpha$ and weight parameters $\omega$. We set the initial learning rate of the optimizer for $\alpha$ to $10^{-3}$, and we set that of the optimizer for $\omega$ to $10^{-2}$. Both learning rate were annealed down to zero following a cosine schedule \cite{loshchilov2016sgdr}. We set the weight decay of both optimizers to $3\times 10^{-4}$. The entire searching process takes around 5 days on a single GPU.

During training from scratch, the model was trained for 300 epochs with a batch size of 128. We explored different initial channels, as we will describe in Section \ref{sec:setup}. We used Adam Optimizer with initial learning rate of $10^{-2}$, which was annealed down to zero following a cosine schedule. We set the weight decay of the optimizer to $3\times 10^{-4}$. The entire training process takes around 1 day on a single GPU.

\subsection{Experimental Setup}
\label{sec:setup}

\begin{figure}
	\centering
	\includegraphics[width=\columnwidth]{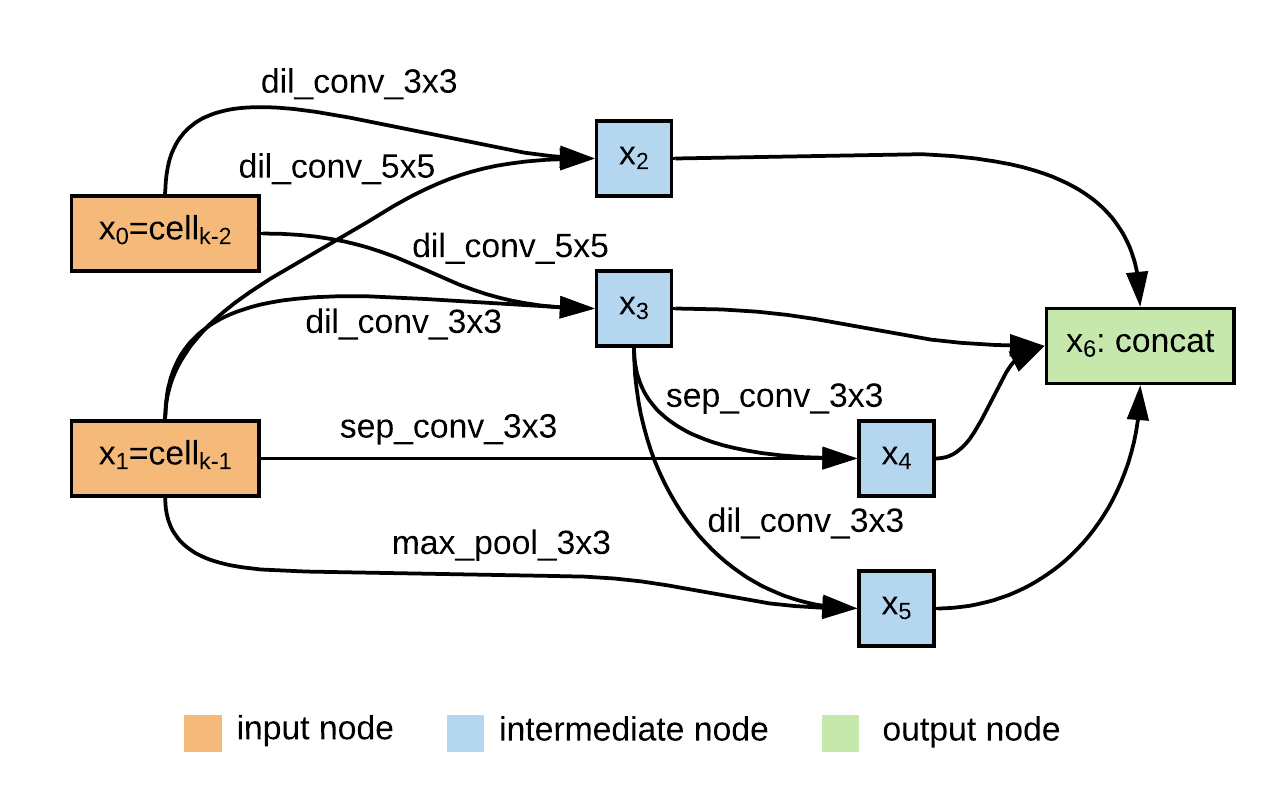}
	\caption{Architecture of the searched normal cell.}
	\vspace{-1em}
	\label{fig:vis_normal}
\end{figure}

We explored the effect of using different number of neural cells and different number of channels during training from scratch. First, we derived a relatively light-weight model by setting the number of neural cells to $N=8$ and the number of initial channels $C=64$. Following this, we tried to construct two larger models by growing the number of neural cells and the number of channels, respectively. For a fair comparison between the two large models, we kept the total number of parameters the same in the two models. Consequently, we have a second model with $N=30, C=64$, and a third model with $N=8, C=128$. The embedding dimensions and the number of parameters of these models are shown in Table \ref{table:results}.

When evaluating the derived architecture for speaker identification, we followed the identification split of VoxCeleb1 dataset. We trained the derived CNN on training set and evaluated it on testing set. To obtain utterance-level speaker embeddings, we divided each test utterance into 3-second segments with 1.5-second overlap. Following this, we extracted the speaker embeddings on these segments, and used the average of these embeddings as the utterance-level speaker embedding. We computed Top-1 and Top-5 accuracies on these utterance-level speaker embeddings to measure the performance.

When evaluating the derived architecture for speaker verification, we followed the verification split of VoxCeleb1 dataset. We trained the derived CNN on training set and evaluated it on the trial pairs. We obtain utterance-level speaker embeddings as described above, and we used cosine similarity as the similarity metric. We used Equal Error Rate (EER) to measure the performance, which is commonly used for speaker verification systems \cite{nagrani2017voxceleb}.

\subsection{Results and Design Insights}

\begin{figure}[h]
	\vspace{0.5em}
	\centering
	\includegraphics[width=\columnwidth]{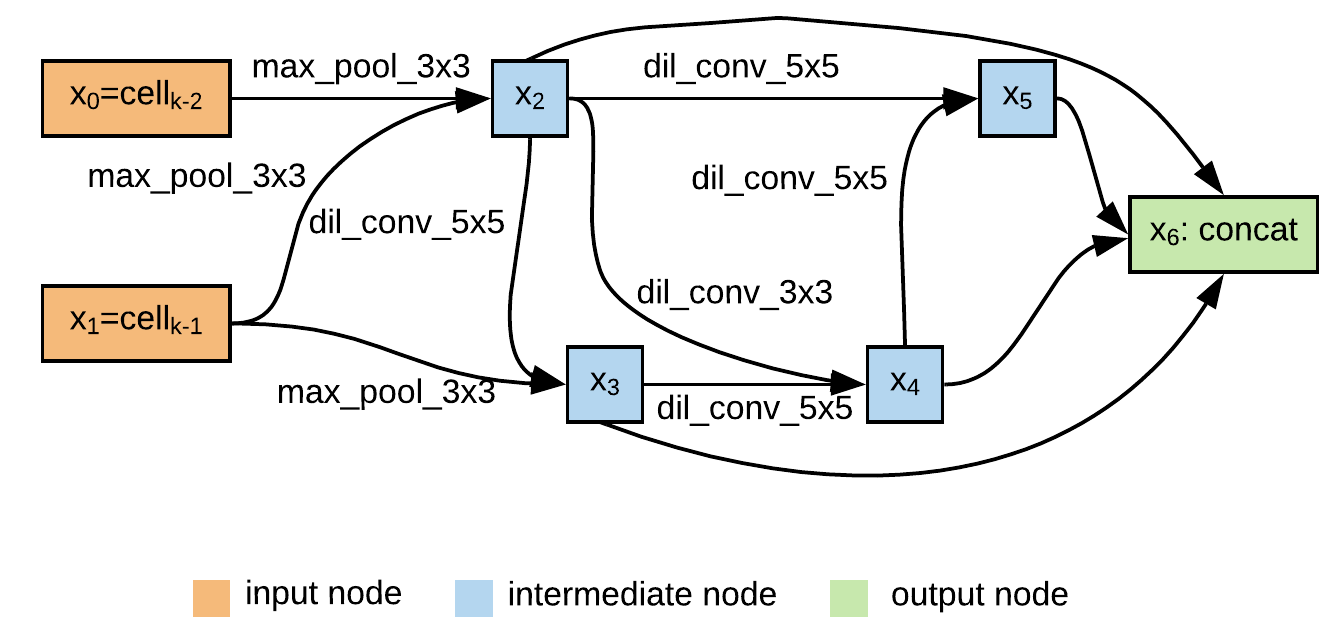}
	\caption{Architecture of the searched reduction cell.}
	\label{fig:vis_reduce}
	\vspace{-1em}
\end{figure}

We visualize the architectures of the searched cells in Figures \ref{fig:vis_normal} and \ref{fig:vis_reduce}. We report the results on speaker identification and speaker verification, as shown in Table \ref{table:results}. 

We first analyze the effect of the number of neural cells and the number of channels in the proposed method. We found that the model with 8 neural cells and 128 initial channels achieved the best performance among the proposed models on both identification (Top-1: $87.66\%$, Top-5: $96.01\%$) and verification (EER: $8.95\%$). Additionally, the two large models both outperform the light-weight one. When comparing between the two large models, we found that the model with ($N=8, C=128$) achieves better performance that the model with ($N=30, C=64$). This result suggests that increasing the number of channels is more effective than increasing the number of neural cells, when constructing a larger model from the searched neural cell.

We next compared the proposed method against three CNN architectures that were commonly used for speaker recognition: VGG-M \cite{nagrani2017voxceleb}, ResNet-18 \cite{hajibabaei2018unified, bhattacharya2019deep}, and RestNet-34 \cite{chung2018voxceleb2, xie2019utterance, cai2018exploring}. As there are no open-source implementations from previous works on ResNet-18 and ResNet-34, we implemented these two architecture under similar training settings as the proposed method for a fair comparison. As shown in Table \ref{table:results}, all our proposed models outperform the baseline CNN architectures. Specifically, our light-weight model ($N=8, C=64$) outperforms ResNet-34, the best baseline architecture, by $3.11\%$ Top-1 and $2.86\%$ EER, with only 5 million parameters (ResNet-34: 22 million parameters). Our best model ($N=8, C=128$) outperforms ResNet-34 by $6.32\%$ Top-1 and $3.04\%$ EER, with only 18 million parameters. These results showed that the searched CNN architectures significantly improved both speaker identification and speaker verification performance, while achieving higher parameter efficiency and enjoying lighter weights.

\section{Conclusions}
\label{sec:conclusions}
This paper proposed an automatic approach to find the optimal CNN architecture for speaker recognition. The proposed approach has two stages. The first stage searches for the optimal architecture of a neural cell consisting of several different types of operations. The second stage derives a CNN model by stacking the searched neural cell several times and then trains the CNN model through the standard speaker recognition training process. Evaluation results on both speaker identification and speaker verification demonstrate significantly improved performance, while having lower model complexity.

\bibliographystyle{IEEEtran}

\bibliography{mybib}


\end{document}